\documentclass[journal]{IEEEtranTIE}
\usepackage{graphicx}
\usepackage{cite}

\usepackage{picinpar}
\usepackage{amsmath}
\usepackage{amssymb}  %
\usepackage{url}
\usepackage{flushend}
\usepackage[latin1]{inputenc}
\usepackage{colortbl}
\usepackage{soul}
\usepackage{multirow}
\usepackage{pifont}
\usepackage{color}
\usepackage{alltt}
\usepackage[hidelinks]{hyperref}
\usepackage{enumerate}
\usepackage{epstopdf}
\usepackage{pbox}
\usepackage[caption=false,font=footnotesize]{subfig}

\definecolor{lightgray}{rgb}{0.3,0.3,0.3}

\begin{document}
\title{Experimental Evaluation of Vibration Influence on a Resonant MEMS Scanning System for Automotive Lidars}

\author{
	\vskip 1em
	Han Woong Yoo, \emph{Membership},
	Rene Riegler, %
	David Brunner, %
	Stephan Albert, \\ %
	Thomas Thurner, \emph{Membership},	
  and Georg Schitter, \emph{Senior Membership}
	\thanks{

		This work has been supported in part by the Austrian Research Promotion Agency (FFG) under the scope of the LiDcAR project (FFG project number 860819).		
		Han Woong Yoo, Rene Riegler, David Brunner, and Georg Schitter are with Automation and Control Institute (ACIN),
        TU Wien, Gusshausstrsse 27-29, 1140, Austria (corresponding to: yoo@acin.tuwien.ac.at). 		
		Stephan Albert is with Infineon Technologies AG, Am Campeon 1-15, 85579, Neubiberg, Germany, and 
		Thomas Thurner is with Design Center Graz, Infineon Technologies AG, Babenbergerstrsse 10, 8020, Graz, Austria.\\
Post-print version of the article: H. W. Yoo, R. Riegler, D. Brunner, S. Albert, T. Thurner, G. Schitter, Experimental
Evaluation of Vibration Influence on a Resonant MEMS Scanning System for Automotive Lidars," IEEE Transactions on 
Industrial Electronics, 2021. DOI: 10.1109/TIE.2021.3065608 \\
\textcopyright 2021 IEEE. Personal use of this material is permitted. Permission from IEEE must be obtained for all other uses, in any
current or future media, including reprinting/republishing this material for advertising or promotional purposes, creating
new collective works, for resale or redistribution to servers or lists, or reuse of any copyrighted component of this work in
other works.
	}
}

\maketitle
	
\begin{abstract}
This paper demonstrates a vibration test for a resonant MEMS scanning system in operation to evaluate the vibration immunity for automotive lidar applications. 
The MEMS mirror has a reinforcement structure on the backside of the mirror, causing vibration coupling by a mismatch between the center of mass and the rotation axis.  %
An analysis of energy variation is proposed, showing  direction dependency of vibration coupling. 
Vibration influences are evaluated by transient vibration response and vibration frequency sweep using a single tone vibration for translational y- and z- axis.  
The measurement results demonstrate standard deviation (STD) amplitude and frequency errors are up to 1.64~\% and 0.26~\%, respectively, for 2~$g_\textrm{rms}$ single tone vibrations on y axis.
The simulation results also show a good agreement with both measurements, proving the proposed vibration coupling mechanism of the MEMS mirror. 
The phased locked loop (PLL) improves the STD amplitude and frequency errors to 0.91~\% and 0.15~\% for y axis vibration, corresponding to 44.4~\% and 43.0~\% reduction, respectively, showing the benefit of a controlled MEMS mirror for reliable automotive MEMS lidars.
\end{abstract}


\markboth{IEEE TRANSACTIONS ON INDUSTRIAL ELECTRONICS}%
{}

\definecolor{limegreen}{rgb}{0.2, 0.8, 0.2}
\definecolor{forestgreen}{rgb}{0.13, 0.55, 0.13}
\definecolor{greenhtml}{rgb}{0.0, 0.5, 0.0}
 
\section{Introduction} %
Lidar, also called ladar or laser radar, allows high accuracy and long range 3D imaging, emerging as an essential sensor technology in high level automated driving systems \cite{royoOverviewLidarImaging2019}.  %
To attain high resolution and long range at the same time, %
MEMS scanners have received much attention as one of the promising beam steering solutions for automotive lidars thanks to their compactness, high reliability with hardly any mechanical wear, long life time, and a low unit price at a large volume production \cite{Yoo18,druml_1d_2018,Kasturi2020Comp}.
Indeed, intensive research and development are ongoing in both industry and academia in the last decade \cite{Ito13,hofmannResonantBiaxial7mm2013,Kim16,Tha16,Sta16,sandnerHybridAssembledMEMS2019}. %

There are, however, several challenges for MEMS mirrors to be a prevailing scanning technique for automotive lidars. 
One of them is robustness in harsh environment conditions, e.g. wide operational climates, varying air pressure and temperatures, excessive vibrations and shocks \cite{wolter_applications_2005-1}. 
Especially for the vibrations, an automotive standard of LV~124 requires a stable operation with a random vibration of at least 30.8~m/s$^2$ RMS acceleration up to 2~kHz \cite{VW8000Std}. 
For destructive tests for MEMS mirrors not in operation, it is aimed by means of design to prevent fracture of the structure due to large accelerations \cite{yooMEMSMicromirrorCharacterization2009}. 
For quasi-static MEMS mirrors, which are non-resonant type and allow wide-band arbitrary beam scanning by special electrostatic actuator designs \cite{VerticalCombs,Jung2012QSFraunhoferIPMS,hofmann_high-q_2012}, a dielectric liquid filling in the mirror package \cite{milanovicNovelFluidicPackaging2016,milanovicIterativeLearningControl2018} or a soft material for fabrication \cite{leiFR4BasedElectromagneticScanning2018} can endure 1000~$g$ shock and 20~$g$ vibrations from 20~Hz to 2~kHz.
For MEMS mirrors in operation, a quasi-static MEMS mirror is evaluated by vibration and shock tests for automotive lidars and parasitic modes are identified as the main cause of vibration coupling to the rotational mode \cite{grahmannVibrationAnalysisMicro2020}. 
For a 10 Hz triangular scan motion of about $\pm$ 4.5$^\circ$ amplitude, a 30~$g$ in-plane acceleration with a frequency of about 500 Hz results in about 0.36$^{\circ}$ peak to peak scanning error for the open loop case and about 0.22$^{\circ}$ for the close loop case, corresponding to about 4.0~\% and 2.4~\% peak to peak scanning error, respectively. 
Furthermore, many conventional MEMS mirror designs have rather small sizes, which limits the resolution of the illuminated spot and reduces the maximum range and SNR of the scanning receiver.
Increasing the size of the MEMS mirror is not straightforward due to dynamic mirror deformation \cite{urey_optical_2000}, causing a blurred light spot in the projection.  
Thicker MEMS mirror can reduce this dynamic deformation, however at the price of a strongly decreased mirror frequency which usually cannot meet the required scanning performance.
One approach employs multiple MEMS mirrors to increase the aperture size, while it demands a highly uniform fabrication and a dedicated synchronization control \cite{sandnerHybridAssembledMEMS2019,sandner_synchronized_2010}.
A popular approach exploits reinforcement structures with a thin mirror to reduce low order surface deformation with a small increase of inertia \cite{neeLightweightOpticallyFlat2000,milanovicGimballessMonolithicSilicon2004,hsuFabricationCharacterizationDynamically2008,hofmannResonantBiaxial7mm2013}. The reinforcement structure can cause a mismatch between the rotational axis and the center of mass of the mirror, leading to coupling to an external vibration.
A recent analysis based on perturbation theory is investigated for a vibration influence of a MEMS mirror with a reinforcement structure \cite{YooCTF2020} while this only studies simulations of an open loop case without feedback and exploits rather impractical input voltages. For resonant MEMS mirrors with reinforcement structures, however, experimental evaluation of the vibration influence during operation has not been reported so far.
The main contribution of this paper is modeling and an experimental evaluation of the vibration influence for a resonant MEMS mirror in operation including a feedback control.
As a main cause of the vibration influence, a mismatch between the center of mass and the rotational axis is raised, which is due to the reinforcement structure of the MEMS mirror. %
To explain the mechanisms of vibration coupling, an energy-based analysis is also proposed. 
A vibration test setup has been developed, which can apply vibration to the MEMS mirror in all three spatial directions while a position sensitive device (PSD) records the scanning trajectories. 
Vibration influences are evaluated by transient vibration response and vibration frequency sweep for an open loop and a phase locked loop (PLL) controlled mirror, using a single tone vibration. 
The measurements are compared to the simulation to verify the proposed vibration coupling model.  %

\section{Description of MEMS Mirror Model and Control}
\begin{figure}[t]
\centering
\centerline{\includegraphics[width=0.47\textwidth]{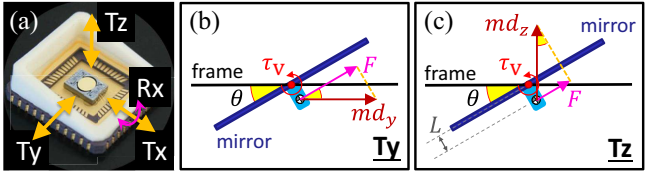}}  
\caption[]{(a) A resonant MEMS scanning mirror for vibration test and the definition of translational axes for vibrations description with respect to the main mirror rotation mode of Rx. (b, c) Vibration coupling torque model by Ty and Tz vibration of $d_y$ and $d_z$ to the Rx motion with the mass $m$ due to the mismatch of $L$  between the center of mass and the rotation axis.} %
\label{fig:vibrationModel}
\end{figure}
Fig.~\ref{fig:vibrationModel}a shows the MEMS scanning mirror used in this paper, equipped with electrostatic comb drive actuators. 
The MEMS mirror has a reinforcement structure on the backside to enable a high scanning frequency with a mirror aperture over 2~mm \cite{leonardusMEMSSCANNINGMICROMIRROR2018}. 
This reinforcement structure shifts the center of mass below the rotational axis, which causes vibration-induced torques from translational y- and/or z-axis vibrations. as illustrated in Fig.~\ref{fig:vibrationModel}b and Fig.~\ref{fig:vibrationModel}c. 

This section firstly describes the dynamic model of the MEMS mirror and the vibration-induced torque model. Then vibration coupling is analyzed based on the energy injection by the vibration-induced torque. Finally, the model of the PI-based PLL is given to describe the PLL-controlled MEMS scanning system.

\subsection{SDoF model of MEMS mirror}
\begin{figure}[t!]
  \centering
  \subfloat[nonlinear stiffness]{\label{fig:stiffness}\includegraphics[width=0.215\textwidth]{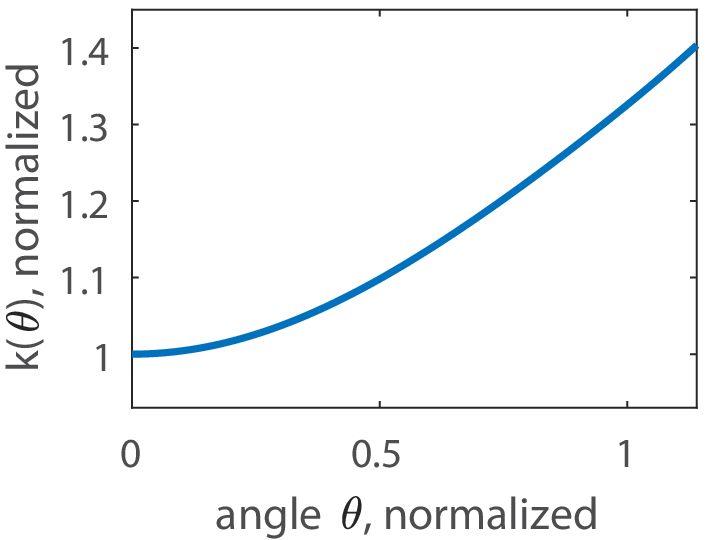}}~
	\subfloat[averaged damping]{\label{fig:damping}\includegraphics[width=0.215\textwidth]{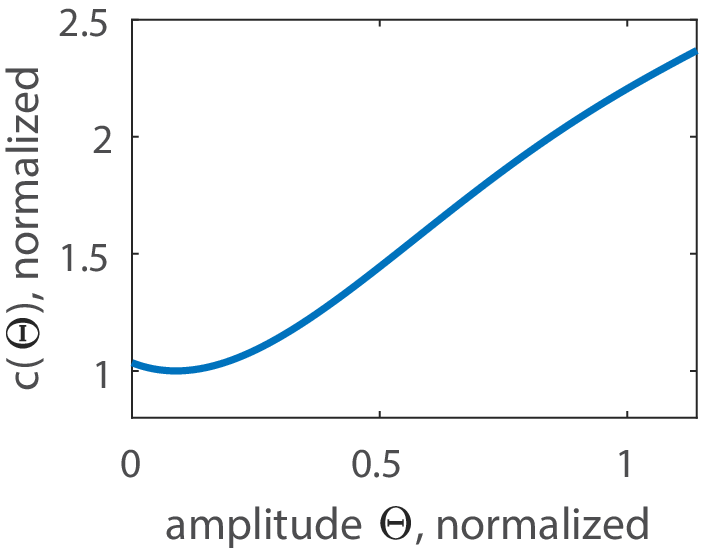}} \\
	\subfloat[comb drive capacitance and its angular derivative]{\label{fig:combCap}\includegraphics[width=0.43\textwidth]{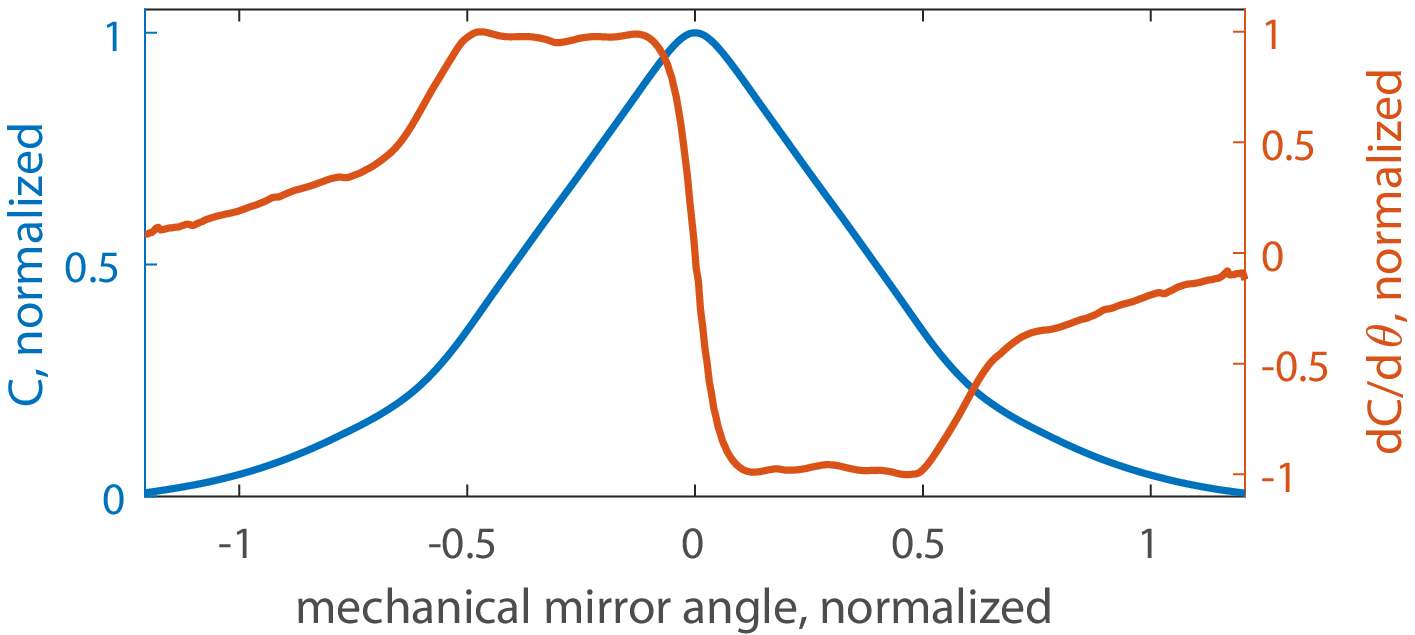}} 
\caption[]{Nonlinear stiffness function, averaged damping function, and capacitance and angular derivative of the capacitance from experimental identification. 
} %
\label{fig:idStiffnessDamping}
\end{figure}
Consider the main rotation mode of the electrostatically actuated MEMS mirror as a single degree of freedom (SDoF) model \cite{DavidSPIE19}
\begin{equation}
	I \ddot \theta + c(\theta,\dot \theta) \dot \theta + k(\theta)\theta  =  \frac{1}{2}\frac{\textrm{d}C}{\textrm{d}\theta} V^2 + \tau_\textrm{v}, \label{eq:ParamDuffOsci}
\end{equation}
where $\theta$ is the mechanical mirror deflection angle in radian, $I$ denotes the moment of inertia of the considered rotational mode, and $c$ and $k$ are nonlinear damping and stiffness functions, respectively. 
The comb drive torque is defined by the product of the squared input voltage function $V$ and the angular derivative of the comb drive capacitance $C$.
$\tau_\textrm{v}$ denotes a torque induced by external vibrations. 
Fig.~\ref{fig:idStiffnessDamping} shows the normalized nonlinear stiffness function, the averaged damping function, and comb drive capacitance, identified from the MEMS mirror.
A detailed discussion of the design concepts and the parameter identification can be found in \cite{YooStephanMEMSDFT20}.

\begin{figure}[t]
    \centering
    \includegraphics[width=0.44\textwidth]{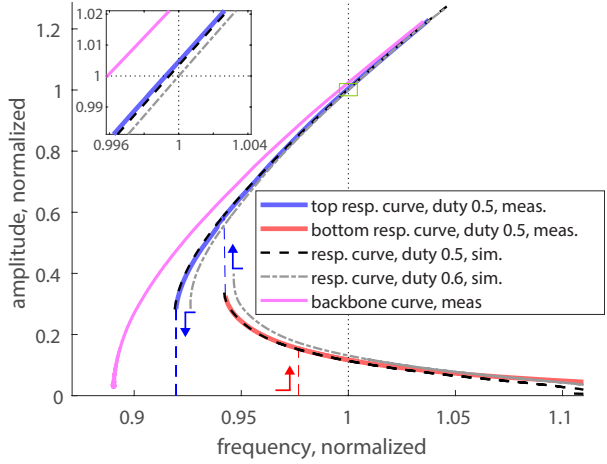}
    \caption[]{Measured and simulated frequency of the MEMS mirror by a 100~V rectangular input including the mirror's backbone curve. The directions of the bifurcation jumps are given by the arrows \cite{DavidSPIE19}. A simulation with a duty of 0.6 is drawn for showing the operation point in the experiments.
} \label{fig:RespCurveMeasSim} %
\end{figure} %
Fig.~\ref{fig:RespCurveMeasSim} shows the frequency response of the MEMS mirror of Fig.~\ref{fig:vibrationModel}. The amplitude and frequency is normalized to the values at the operation point in the experiment. 
The MEMS mirror is excited to the 1st order parametric resonance by the electrostatic comb drive, i.e. the actuation frequency is twice the mirror frequency \cite{Ataman2006,frangiParametricResonanceElectrostatically2017,frangiAccurateSimulationParametrically2017,kovacicMathieuEquationIts2018}. The backbone curve and the frequency response illustrate bending toward higher frequencies due to the nonlinear hardening of the chosen suspension structure.
This suspension structure is beneficial to increase scanning frequency and suppress other unwanted mode of the mirror motion. 
The identification algorithm and ODE simulation of (\ref{eq:ParamDuffOsci}) without external torques are developed and verified in \cite{DavidSPIE19}, showing a good agreement with the measurement data. 

\subsection{External vibration-induced torque model} \label{se:energyAnalysis}
Consider accelerations of $d_{y}$ and $d_{z}$ caused by a translational vibration along the y- and z-axis in Fig.~\ref{fig:vibrationModel}, which are called hereafter Ty vibration and Tz vibration. Due to the mismatch of the center of mass and the rotational axis, the vibration-induced torque is generated as  %
\begin{equation}
 \tau_\textrm{v} =  L m d_{y} \cos \theta + L m d_{z} \sin \theta, \label{eq:DuffWithNonColDist} 
\end{equation}
where $m$ and $L$ denote the mass of the mirror and the distance between the rotational axis and the center of mass, respectively. The vibration-induced torque is scaled by the cosine or sine of the mirror angle $\theta$ due to the movement of the center of mass. For simplicity in analysis, only a single tone vibration is  considered as  
\begin{align}
d_{y} = a_{y} \cos(2 \pi f_y t + \phi_y),  \label{eq:singleToneTy} \\ 
d_{z} = a_{z} \cos(2 \pi f_z t + \phi_z), \label{eq:singleToneTz} 
\end{align} %
where $a_{y}$ and $a_{z}$ are the amplitude of vibration with the frequency, $f_y$ and $f_z$, and the phase to the mirror, $\phi_y$ and $\phi_z$, for the Ty and Tz vibration, respectively. Due to the nonlinear dynamics of (\ref{eq:ParamDuffOsci}), superposition of the vibration-induced torque does not typically hold for a large amplitude vibration. For a small vibration around a stable equilibrium point, i.e. steady state operation of the mirror,  (\ref{eq:ParamDuffOsci}) can be linearized, leading to a generalization by the superposition of the single tone analysis. The conditions of linearization are not discussed further in this paper nevertheless the linearization is used for the analysis of the vibration coupling to the mirror motion. 

\subsection{External vibration coupling model based on energy variation} %
Consider the MEMS mirror to be in a steady state, i.e. the energy gain and energy loss from injections by the comb drives and damping are balanced. 
In this steady state, the external torque brings a change of the energy in the mirror motion, leading to a variation of amplitude and frequencies. 
Assume that the mirror trajectory is approximated by a single tone sine with a steady state mirror frequency $f_\mathrm{m}$ and an amplitude $\Theta$. The errors of the frequency and amplitude by the vibration are much smaller than the steady state frequency and amplitude, i.e. $\Delta f_\textrm{m} \ll f_\textrm{m}$ and $\Delta \Theta \ll \Theta$. 
Then the mirror angle can be approximated as 
\begin{equation}
\theta \approx \Theta \sin 2 \pi f_\textrm{m} t. \label{eq:singletoneApprox}
\end{equation}
Consider 20$^\circ$ as a maximum deflection angle, i.e. $\Theta<0.35$ in radian. Taylor approximation can expand the single tone vibration-induced torque of (\ref{eq:DuffWithNonColDist}) with (\ref{eq:singleToneTz}) and (\ref{eq:singleToneTy}) as
\begin{align}
 \tau_\textrm{v} &\approx   m L a_{y} \cos(2 \pi f_y t + \phi_y) (1 -\frac{1}{2} \theta^2) \nonumber \\
					&+  m L a_{z} \cos(2 \pi f_z t + \phi_z) (\theta - \frac{1}{6} \theta^3). \label{eq:vibrationTorqueApprox1}
\end{align}
Besides, the vibration-induced energy change at $t$ for a single period of (\ref{eq:singletoneApprox}) can be written by an average as 
\begin{align}
\Delta E_{\textrm{v},t} &= \int^{t+ \frac{1}{f_\textrm{m}}}_{t} \tau_\textrm{v}(\eta) \dot \theta(\eta) \textrm{d}\eta, \label{eq:DeltaEnergyBasic}
\end{align}
where $\eta$ is the integration variable of time. Assume that $f_y \neq (2 n-1) f_\textrm{m}$ and $f_z \neq 2 n f_\textrm{m}$ for $n = 1, 2$.
With trigonometric identities,  
substituting  (\ref{eq:singletoneApprox}) and (\ref{eq:vibrationTorqueApprox1}) into (\ref{eq:DeltaEnergyBasic})  leads to 
\begin{align}
\Delta E_{\textrm{v},t}	   		 &\approx + \left. a_{y} v_{y,1} f_\textrm{m} \frac{\sin (2 \pi ( f_\textrm{m} - f_y) \eta - \phi_y)}{2 \pi (f_\textrm{m} - f_y)}  \right  |^{\eta =t+\frac{1}{f_\textrm{m}}}_{\eta =t}    \nonumber \\
                 & + \left. a_{y} v_{y,3} f_\textrm{m} \frac{\sin (2 \pi ( 3 f_\textrm{m} - f_y) \eta - \phi_y)}{2 \pi (3 f_\textrm{m} - f_y)}  \right  |^{\eta =t+\frac{1}{f_\textrm{m}}}_{\eta =t}      \nonumber \\								
								& - \left. a_{z} v_{z,2} f_\textrm{m} \frac{\cos (2 \pi (2 f_\textrm{m} - f_z) \eta - \phi_z)}{2 \pi (2 f_\textrm{m} - f_z)} \right |^{\eta =t+\frac{1}{f_\textrm{m}}}_{\eta =t}  \nonumber \\
								 & - \left. a_{z} v_{z,4} f_\textrm{m} \frac{\cos (2 \pi ( 4 f_\textrm{m} - f_z) \eta - \phi_z)}{2 \pi (4 f_\textrm{m} - f_z)} \right  |^{\eta =t+\frac{1}{f_\textrm{m}}}_{\eta =t}, \label{eq:DeltaEnergyBeating}
\end{align}
\normalsize
where the vibration coupling coefficients are defined by
\begin{align*}
&v_{y,1} =  \frac{v_0}{16} (8 - \Theta^2), \quad  v_{y,3} =  \frac{v_0}{16} \Theta^2,  \\ 
&v_{z,2} =  \frac{v_0}{48}( 12 \Theta - \Theta^3 ), \quad v_{z,4} =  \frac{v_0}{96} \Theta^3, \quad  v_0 = 2 \pi  m L  \Theta.
\end{align*}
For the approximation, the terms with a higher frequency than $f_\textrm{m}$, e.g. $f_\textrm{m} + f_y$, are omitted since they are small by the averaging integral. 
Equation (\ref{eq:DeltaEnergyBeating}) indicates that the vibration coupling results in energy changes varying at the differences between mirror and vibration frequencies and that the orders of harmonics are different for the two directions.
At the considered amplitudes of $\Theta < 0.35$, the respective first order contributions are dominant, i.e. $v_{z,2} \gg v_{z,4}$ and $v_{y,1} \gg v_{y,3}$.

Ty vibration near $f_\textrm{m}$ and Tz vibration near $2f_\textrm{m}$ are mainly considered, i.e. $2 f_\textrm{m} - f_z \ll f_\textrm{m}$ and $f_\textrm{m} - f_y \ll f_\textrm{m}$ since the local dynamics of amplitude and frequency at an equilibrium typically has a much lower bandwidth than the mirror frequency.  
This allows further approximation of (\ref{eq:DeltaEnergyBeating}) as
\begin{align}
\Delta E_{\textrm{v},t} &\approx   a_{y} v_{y,1} \cos \left (2 \pi \left ( f_\textrm{m} - f_y \right ) t - \phi_y \right ) \nonumber \\
&+  a_{z} v_{z,2}  \sin \left (2 \pi \left (2 f_\textrm{m} - f_z \right  ) t - \phi_z \right ).  \label{eq:vibInflEnergy}
\end{align}
This result implies four aspects of vibration coupling to the scanning motion of the MEMS mirror in operation. 
First, the injected energy per period by vibrations can be approximated by a sinusoidal function with the frequency difference between the vibration frequency and the mirror frequency or the mirror actuation frequency. %
Second, vibrations near the mirror frequency or the mirror actuation frequency are only coupled to the mirror dynamics, hence representing band-limited local dynamics at an equilibrium.
Third, vibration sensitivity with respect to vibration frequencies depends on the direction of the vibration, e.g. high sensitivity for Ty vibration with frequencies near the mirror frequency and high sensitivity for Tz vibration with frequencies near mirror actuation frequencies. 
Last, coupling of Ty vibrations to the mirror dynamics is expected to be stronger than coupling of Tz vibrations, considering $\Theta < 0.35$. 
These four aspects are discussed further below with simulations and measurement results.  

\subsection{PI-based phase locked loop}
\begin{figure}   
    \centering
    \includegraphics[width=0.49\textwidth]{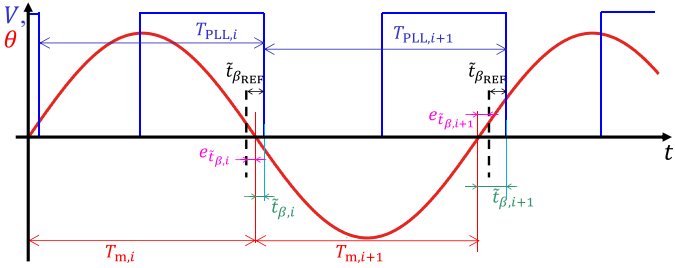}
    \caption[]{Definitions of a PLL operation on the top response curve: the period of PLL $T_\textrm{PLL}$, the half period mirror $T_\textrm{m}$, reference phase in time $\tilde t_{\beta_\textrm{REF}}$, actual phase in time $\tilde t_{\beta}$, and phase error $e_{\tilde t_{\beta}}$ in time based on the mirror angle and the actuation signal, where $i$ denotes the index of the PLL periods} \label{fig:PLLdef}
\end{figure}
To describe the used PI-based PLL, a period-based frequency and a phase in time are considered as depicted in Fig.~\ref{fig:PLLdef}.
The evolution of the phase between the mirror angle and the input voltage in time $\tilde t_{\beta,i}$ is defined by \cite{DavidTIE20}
\begin{equation}
\tilde t_{\beta,i+1} = \tilde t_{\beta,i} - T_{\textrm{m},i+1} + T_{\textrm{PLL},i+1}, \label{eq:PeriodEq}
\end{equation}
where  $T_{\textrm{PLL},i}$ denotes the $i$-th period of the PLL, $T_{m,i}$ denotes the $i$-th half period of the mirror, and where $i$ denotes the index of the PLL periods. Half the mirror period is used instead of the full since the PLL frequency is twice the mirror frequency in steady state. %
The PLL aims to keep the phase to the reference phase, interpreted as a time delay $\tilde t_{\beta_\textrm{REF}}$. 
The phase error is defined by the phase error to the reference phase as $e_{\tilde t_{\beta,i}} = \tilde t_{\beta_\textrm{REF}}- \tilde t_{\beta,i}$. 
A PI controller can be defined as %
\begin{equation} 
 T_{\textrm{PLL},i+1} =  k_\textrm{P} e_{\tilde t_{\beta,i}} +  k_\textrm{I} \sum^{j = i}_{j=0} e_{\tilde t_{\beta,i}} + T_{\textrm{PLL},0}, \label{eq:PLLEqT0}
\end{equation} %
where $k_\textrm{P}$  and $k_\textrm{I}$ are P and I gain, respectively. 
The evolution form is obtained by taking the difference between $i+1$ and $i$ as \cite{DavidTIE20}
\begin{align}
T_{\textrm{PLL},i+1} &= T_{\textrm{PLL},i} + k_\textrm{P} \left (e_{\tilde t_{\beta,i}} - e_{\tilde t_{\beta,i-1}} \right)  +  k_\textrm{I} e_{\tilde t_{\beta,i}} \nonumber \\
			      &= T_{\textrm{PLL},i} + k_\textrm{P} \left (T_{\textrm{m},i} - T_{\textrm{PLL},i} \right)  +  k_\textrm{I} e_{\tilde t_{\beta,i}}.
\end{align} %
This illustrates that the designed PI-based PLL compensates the errors in phase and frequency by I and P gain, respectively. 
Therefore roles of P and I gain in the control are different and both gains are necessary to converge fast to the target phase. 
A detailed analysis of the PI-based PLL is discussed for linearized local dynamics in \cite{DavidTIE20}. %

\section{Experimental Setup and Implementation}
\subsection{Vibration test setup}
\begin{figure}%
    \centering
    \includegraphics[width=0.40\textwidth]{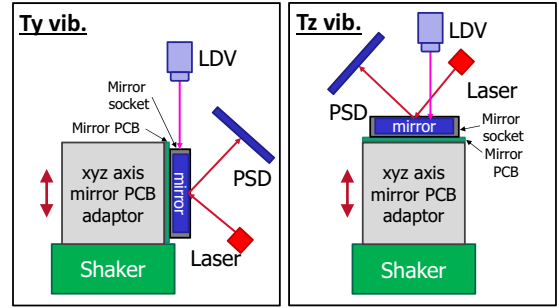}
    \caption[]{Schematics of the vibration test setup for Ty and Tz vibration. A PCB adapter cube allows the installation of the mirror PCB for Ty and Tz vibrations. The acceleration of the MEMS mirror is measured by a laser Doppler vibrometer (LDV) while the mirror scanning trajectory is recorded by a PSD. } \label{fig:ExperimentalSetup} 
\end{figure}
Fig.~\ref{fig:ExperimentalSetup} illustrates a vibration test setup for evaluation of vibration influences on a MEMS scanning system. 
A shaker (TV 51110-M, Tira GmbH, Schalkau, Germany) generates a single directional vibration to a PCB adapter cube.
The PCB adapter cube delivers the vibration to the three possible mirror PCB locations, allowing for Tx, Ty, and Tz vibration test. 
The mirror PCB is tightly attached to one of the faces of the cube without a gap so that unwanted modes induced by the PCB are suppressed. 
While the vibration is applied to the MEMS mirror, the mirror trajectory is measured with a 1D PSD (1L30\_SU2, SiTek Electro Optics, Partille, Sweden), using a collimated fiber laser (S1FC635 with CFC-5X-A, Thorlabs, Newton, NJ, USA). The cross-coupling between vibrations and the PSD angle measurement is negligible compared to the vibration influence on the mirror angle trajectory.

For the control of the vibration and the data acquisition of the measured PSD signal, an FPGA module in a PXIe system (NI PXIe 7856-R, Austin, TX, USA) is used. 
The velocity of the vibration is measured by a laser Doppler vibrometer (OFV 534 with OFV 5000, Polytec GmbH, Waldbronn, Germany). 
From the measured velocity, the amplitude of the single tone sine vibration is determined to keep a constant target acceleration over various vibration frequencies. 
For Tz vibration, the vibrometer measures directly the frame of the MEMS mirror while the socket is measured instead for Ty vibration since direct measurement of the MEMS mirror is not applicable. 

A model based calibration is used for the accurate conversion from the beam position on the PSD to the mirror angle. 
Contrary to the MEMS test bench \cite{YooTestBench2019}, a stage-based calibration scheme is not possible because high stiffness of the PSD installation for robust angle measurements is essential.
Therefore, the mirror parameters in (\ref{eq:ParamDuffOsci}) are identified with the MEMS test bench in advance and the operational distance between the MEMS mirror and the PSD is calculated by matching the measurement with the simulated amplitude. 

\subsection{Implementation of the mirror control}
The mirror is operated by a control ASIC board, an FPGA implementation of the MEMS Driver ASIC for feasibility studies \cite{Yoo18,druml_1d_2018}. 
The board is capable of both open loop operation and PI-based PLL operation.  %
The mirror frequency is detected via the zero crossing of the comb drive currents. A detailed description of the PLL implementation can be found in \cite{druml_1d_2018,Ievgeniia2019SPIE}. %

\subsection{Numerical implementation of simulator}
The ODE simulator for vibration influence is an extension from the SDoF simulation platform of the MEMS mirror in \cite{DavidSPIE19}. 
The vibration torque is added to the previous model and a PLL model is implemented based on the behavioral model of the FPGA implementation. 
The mirror dynamics (\ref{eq:ParamDuffOsci}) are implemented by compiled S-function in Matlab Simulink, significantly reducing computation time. %

\section{Evaluation Results of Vibration Influence}
\begin{figure}
	\centering
	\includegraphics[width=0.49\textwidth]{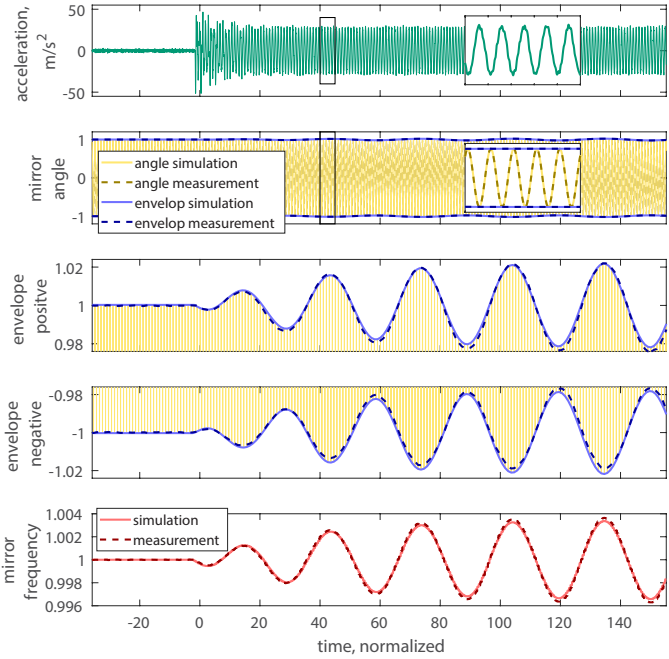}
	\caption[Open Loop]
{Transient response of scanning trajectory, envelope for amplitude changes, and frequency changes of the open loop mirror system for a 2~$g_\textrm{rms}$ Ty vibration with a frequency of 1.0327. %
}\label{fig:TyTrans}
\end{figure}
Vibration influences are evaluated by two experiments: transient vibration response and vibration frequency sweep. 
transient vibration response evaluates a response  of the mirror angle in time to a single tone vibration with a step amplitude. 
The vibration frequency sweep measures steady state vibration responses in mirror amplitude and frequency to a single tone vibration with a normalized frequency from 0.42 to 2.09, providing the sensitivity of the scanning motion to a specific vibration frequency. 
For both experiments, the acceleration of the single tone vibration is set to 2~$g_\textrm{rms}$, corresponding to a peak to peak acceleration of 55.48~$\textrm{m/s}^2$. 
 The strong single tone acceleration of 2~$g_\textrm{rms}$ is chosen to attain a reasonable SNR to characterize the vibration influence by Tz vibrations. 
This single tone excitation is considered much harsher than expected in automotive applications, as demanded by automotive standards, e.g. the LV124  \cite{VW8000Std} with wide-band vibration of in total 11.83~$\textrm{m/s}^2$ RMS spread over the frequency range from 1~kHz to 2~kHz.
 In addition, the mirror frequency is higher than 2~kHz by design, where vibration influence is not defined by the test standard. %
As in Fig.~\ref{fig:RespCurveMeasSim}, amplitudes and frequencies are normalized to the values at the operation point, i.e. the mirror is operated at the frequency $1$ with an amplitude of $1$. %
The open loop MEMS scanning system is evaluated first and then the PLL-controlled MEMS scanning system is examined. 
\subsection{Open loop MEMS scanning system}
\begin{figure}
    \centering
    \includegraphics[width=0.49\textwidth]{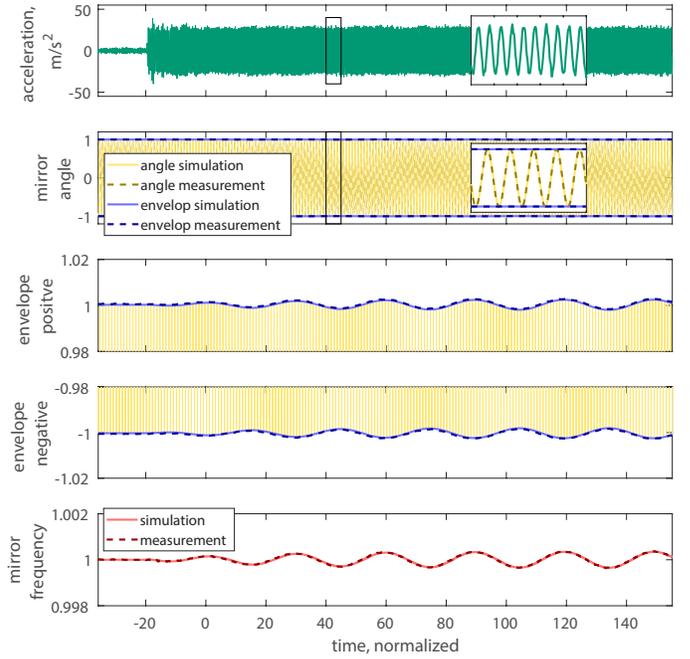}
  \caption[Open Loop]
{Transient response of scanning trajectory, envelope for amplitude changes, and frequency changes of the open loop mirror system for a 2~$g_\textrm{rms}$ Tz vibration with
a frequency of 2.0331. %
}\label{fig:TzTrans}
\end{figure}
The open loop MEMS scanning system is evaluated first because it shows vibration coupling to the pure MEMS mirror dynamics. 
In the open loop case, the MEMS mirror is operated with the duty cycle of 0.6 as shown in Fig~\ref{fig:RespCurveMeasSim}. %

Fig.~\ref{fig:TyTrans} illustrates the transient response of a Ty vibration with a frequency of 1.0327. %
When the vibration starts, the amplitude of the scanning trajectory oscillates at a frequency of 0.0327, corresponding to the frequency difference between the vibration frequency and the mirror frequency. %
The envelope of the positive and negative amplitude shows that the mean amplitude stays the same but the amplitude oscillates with a peak to peak amplitude error of 0.0472. 
The mirror frequency also oscillates at the frequency difference with peak to peak frequency error of 0.0068. 
The simulation is conducted by the measured vibration as a vibration input, showing a good agreement with the measurements data.

Fig.~\ref{fig:TzTrans} shows a transient response for a Tz vibration with a frequency of  2.0331. 
The influence on both the mirror amplitude and frequency are a single tone oscillation with a frequency of 0.0331, which is the difference of the vibration frequency from the mirror actuation frequency. 
The peak to peak errors of the mirror amplitude and frequency are 0.0044 and 0.0007, respectively, which is about a factor of 10 less than those with Ty vibration.

\begin{figure}%
  \centering
  \subfloat[amplitude]{\label{fig:TyamplitudeS}\includegraphics[width=0.5\textwidth]{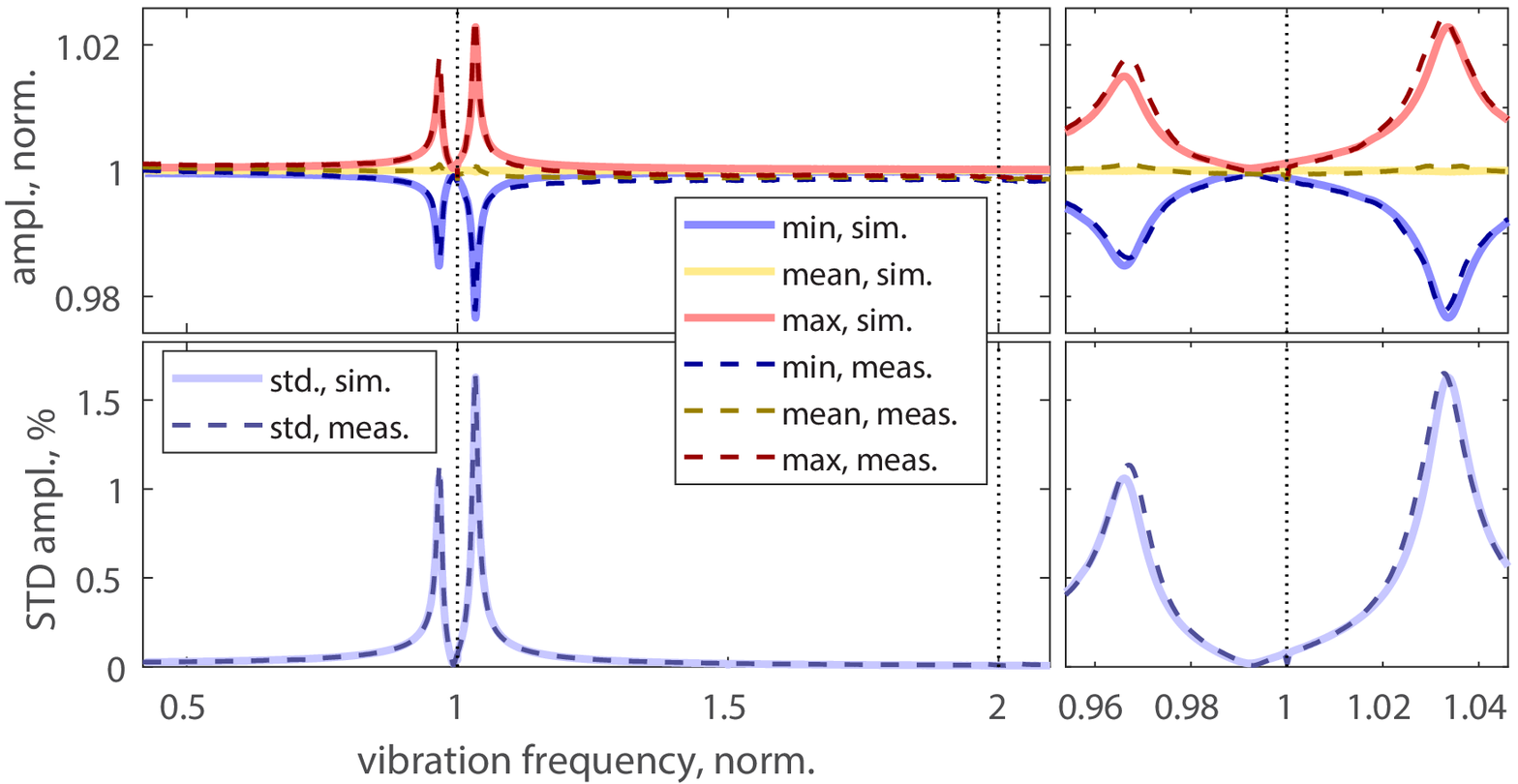}} \\
	\subfloat[frequency]{\label{fig:TyfrequencyS}\includegraphics[width=0.5\textwidth]{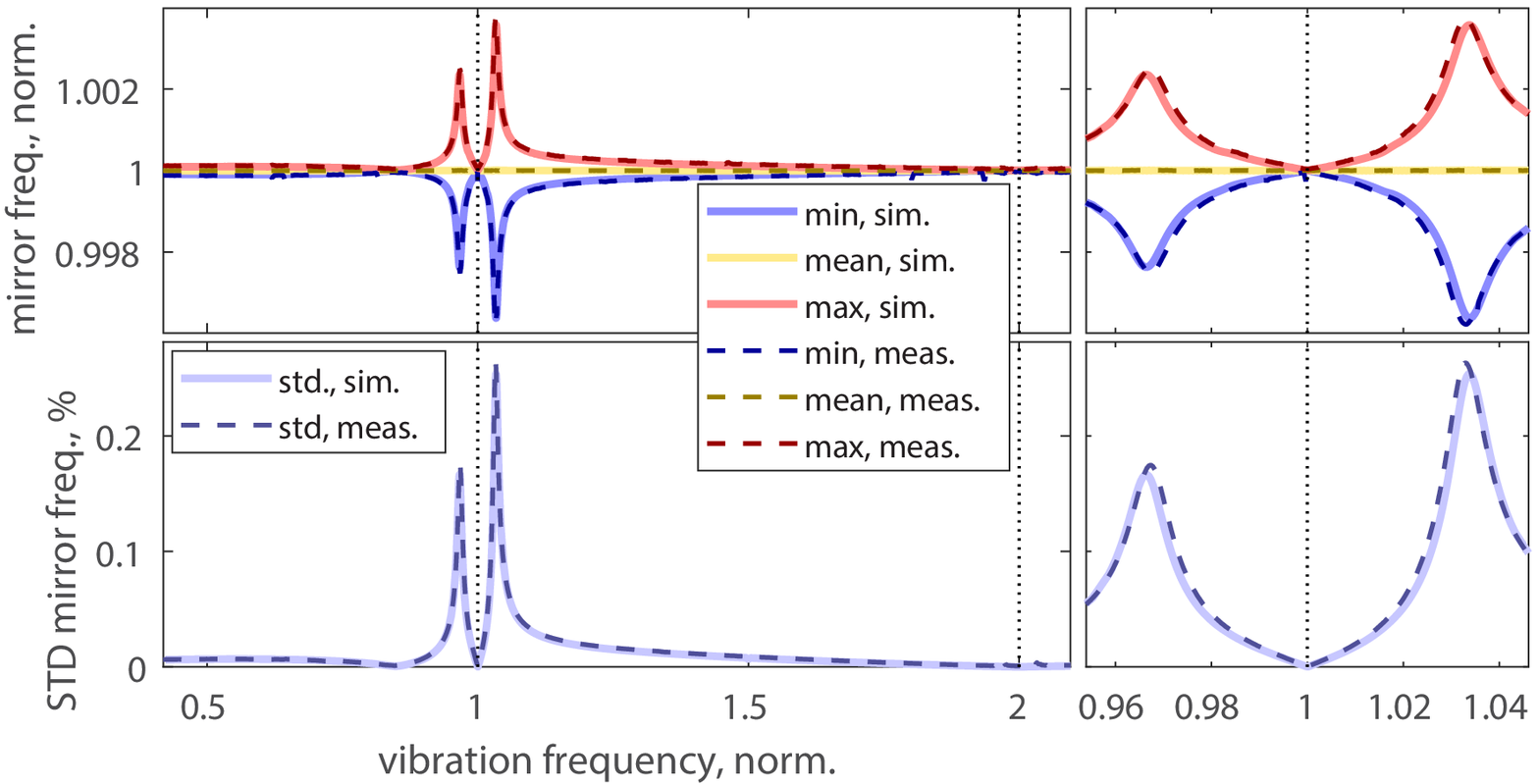}} 
	\caption[]{Ty vibration influence on the mirror amplitude and frequency versus the vibration frequencies in open loop operation. The simulation and measurement are drawn with light solid lines and dark dashed lines, respectively. The influence is shown in the variation of amplitude, represented by the maximum, mean, and minimum amplitude during the vibration. The standard deviation is also drawn to show the results with less noise influence. } 
\label{fig:TySweep}
\end{figure}   
\begin{figure}%
  \centering
  \subfloat[amplitude]{\label{fig:TzamplitudeS}\includegraphics[width=0.5\textwidth]{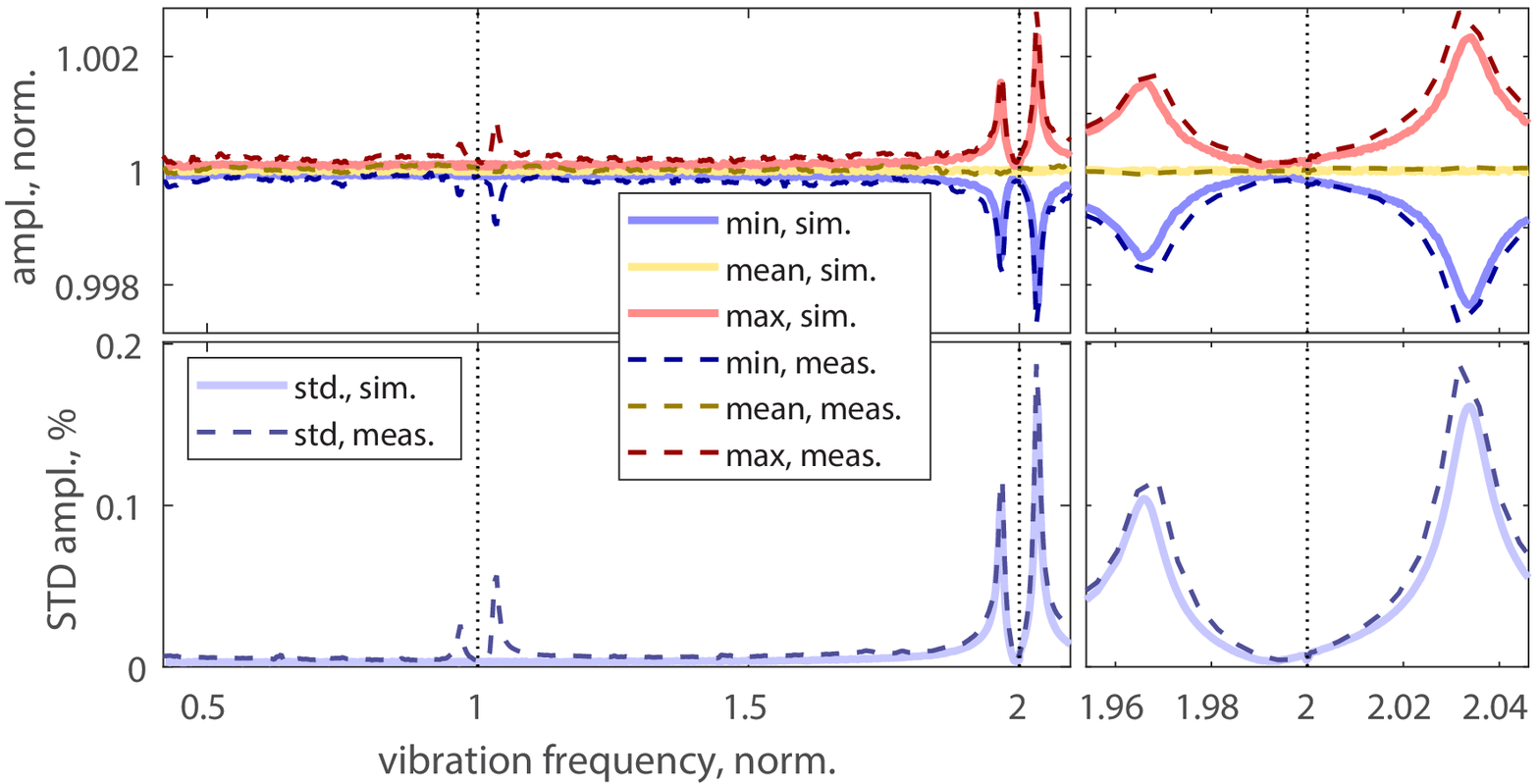}} \\
	\subfloat[frequency]{\label{fig:TzfrequencyS}\includegraphics[width=0.5\textwidth]{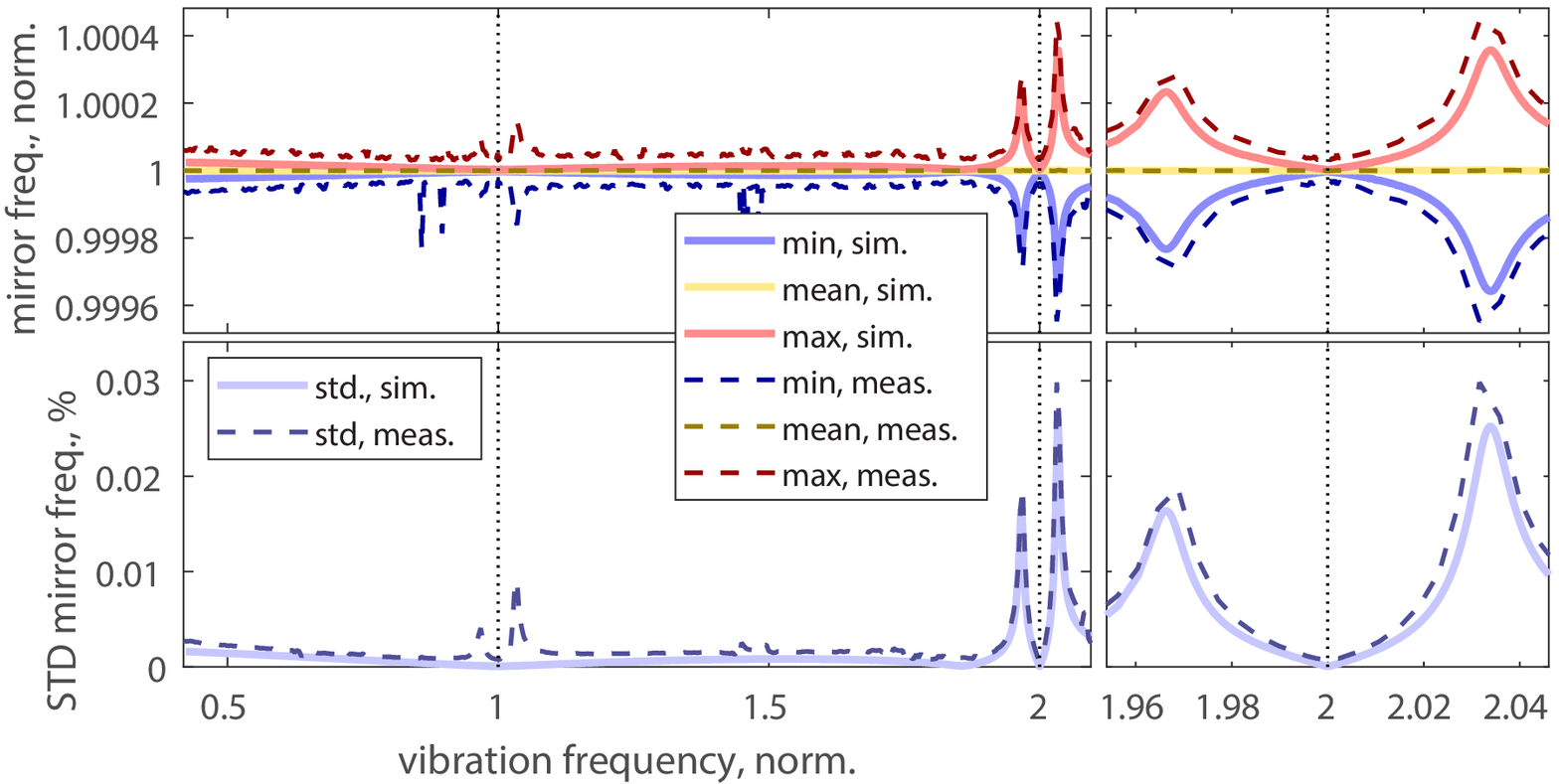}} 
	\caption[]{Tz vibration influence on the mirror amplitude and  frequency for a frequency sweep in open loop operation.  The simulation and measurement are drawn with light solid lines  and dark dashed lines, respectively. } 
\label{fig:TzSweep}
\end{figure}   

Fig.~\ref{fig:TySweep} illustrates a vibration frequency sweep with Ty vibrations. 
The mean of amplitude and frequency tends to stay constant for all vibration frequencies while the maximum and the minimum amplitudes and frequencies vary symmetrically with respect to the mean.
Vibration coupling is spread over both positive and negative frequency differences near the mirror frequency and the peaks are at $\pm 0.033$.
Amplitude and frequency errors can also be evaluated by standard deviation (STD), having the advantage of being a robust measure to noise. 
The STD amplitude errors are 1.13~\% and 1.64~\% at the peaks for the negative and positive frequency differences, respectively, %
and the STD frequency errors are 0.17~\% and 0.26~\% at the peaks for the negative and positive frequency differences, respectively. 
The amplitude and frequency errors at the negative frequency difference are about 31.1~\% and 34.6~\% lower than those at positive frequency difference, respectively.
The simulation results of (\ref{eq:ParamDuffOsci}) also show peaks at $\pm 0.034$ from the mirror frequency, demonstrating a good agreement with measurements. 
In both measurements and simulations, a zero influence in the STD amplitude error is observed at -0.008 from the mirror frequency. %
For the STD frequency error, the zero influences are located at the mirror frequency at -0.154 from the mirror frequency. 

\begin{figure} [t!]
    \centering
    \includegraphics[width=0.49\textwidth]{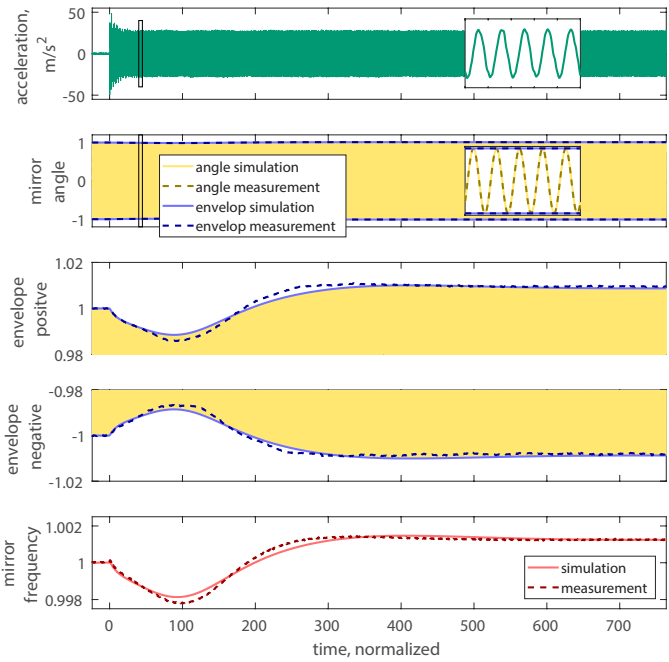}
    \caption[PLL]
{Transient response of scanning trajectory, envelope for amplitude changes, and frequency changes of the PLL-controlled mirror system for a 2~$g_\textrm{rms}$ Ty vibration with a frequency of 1.0013. %
} \label{fig:TyTransientPLL}
\end{figure}

Fig.~\ref{fig:TzSweep} shows the vibration frequency sweep for Tz vibration.
The shape of the vibration influence is similar to that of Ty vibration, e.g. the location of the peak and zeros in STD amplitude and frequency errors. 
The main differences are that the vibration influences with Tz vibration are near the mirror actuation frequency, and are about 10 times smaller than the Ty vibration case with a peak STD amplitude and frequency error of 0.20~\% and 0.03~\%, respectively. 
A small vibration influence near the mirror frequency is observed, which is due to a Ty vibration component caused by a small angle error of about 2$^\circ$ between the Tz vibration and the z direction of the MEMS mirror at the zero angle.

\subsection{PLL-controlled MEMS scanning system}
\begin{figure} [t!]
  \centering
  \subfloat[amplitude]{\label{fig:TyamplitudePLL}\includegraphics[width=0.5\textwidth]{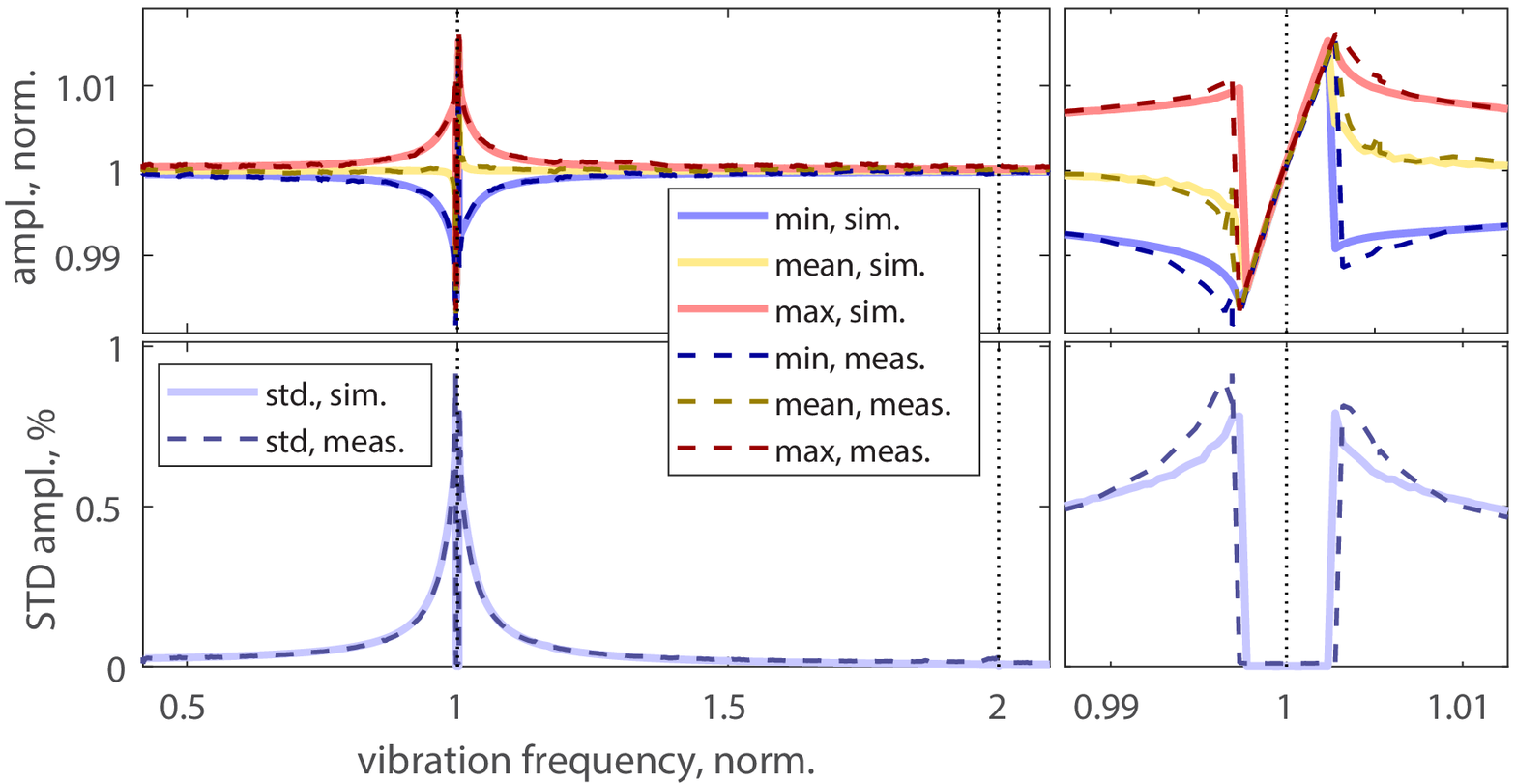}} \\
	\subfloat[frequency]{\label{fig:TyfrequencyPLL}\includegraphics[width=0.5\textwidth]{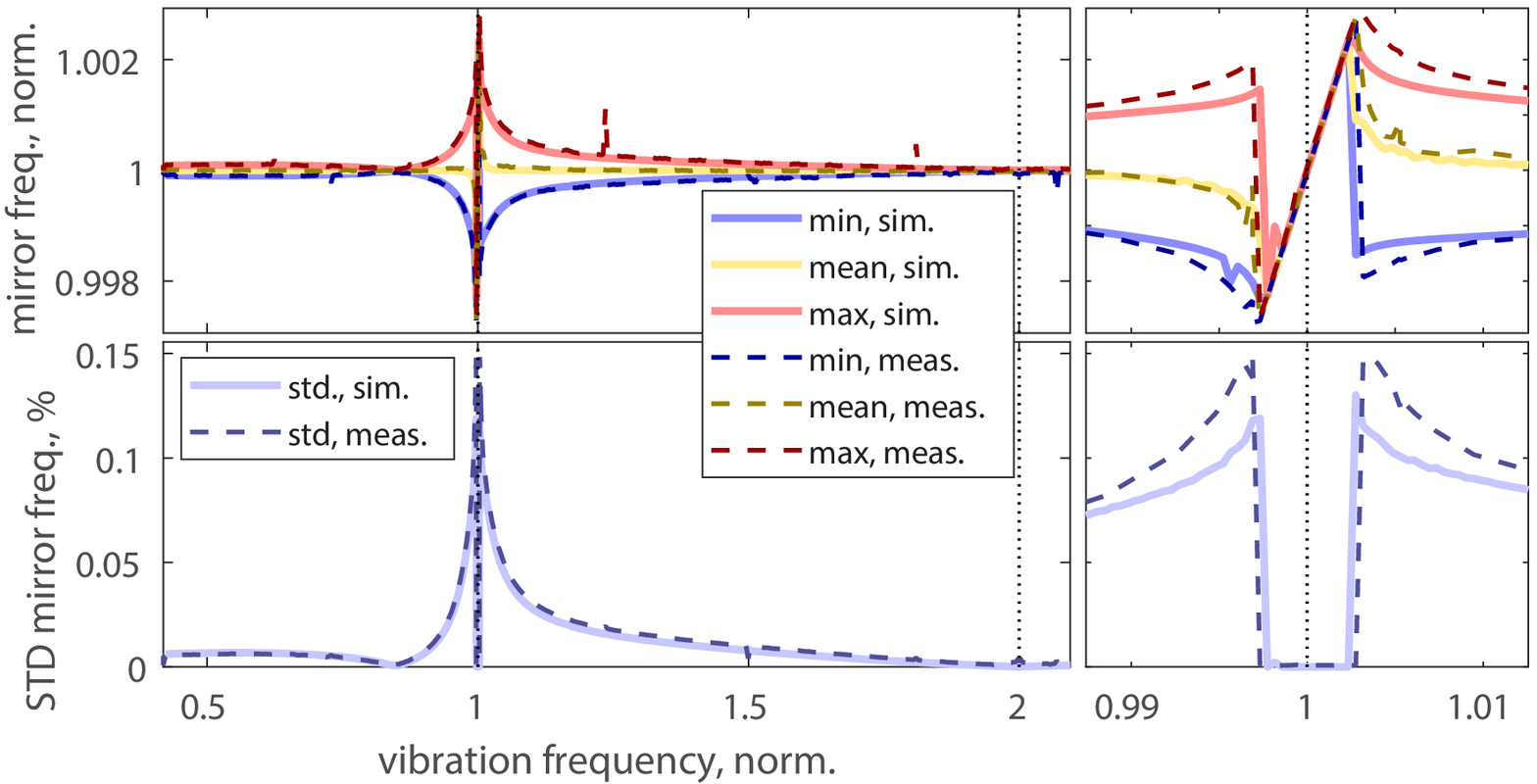}} 
	\caption[]{Ty vibration influence on the amplitude and  frequency for the MEMS mirror with a PLL. The simulation and measurement are drawn with solid lines with light colors and dashed lines with dark colors, respectively. } 
\label{fig:TySweepPLL}
\end{figure}
\begin{figure}%
  \centering
  \subfloat[amplitude]{\label{fig:TzamplitudePLL}\includegraphics[width=0.5\textwidth]{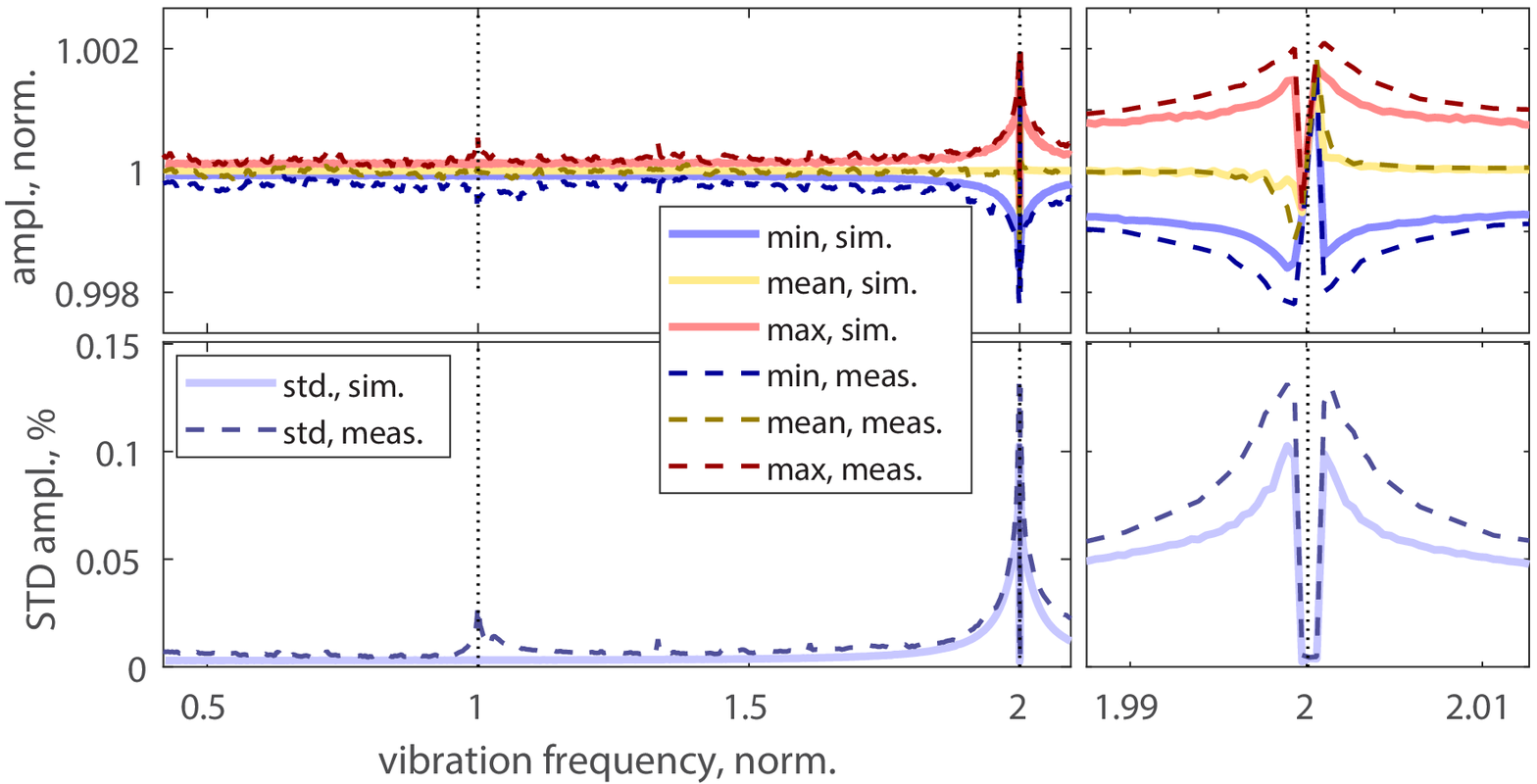}} \\
	\subfloat[frequency]{\label{fig:TzfrequencyPLL}\includegraphics[width=0.5\textwidth]{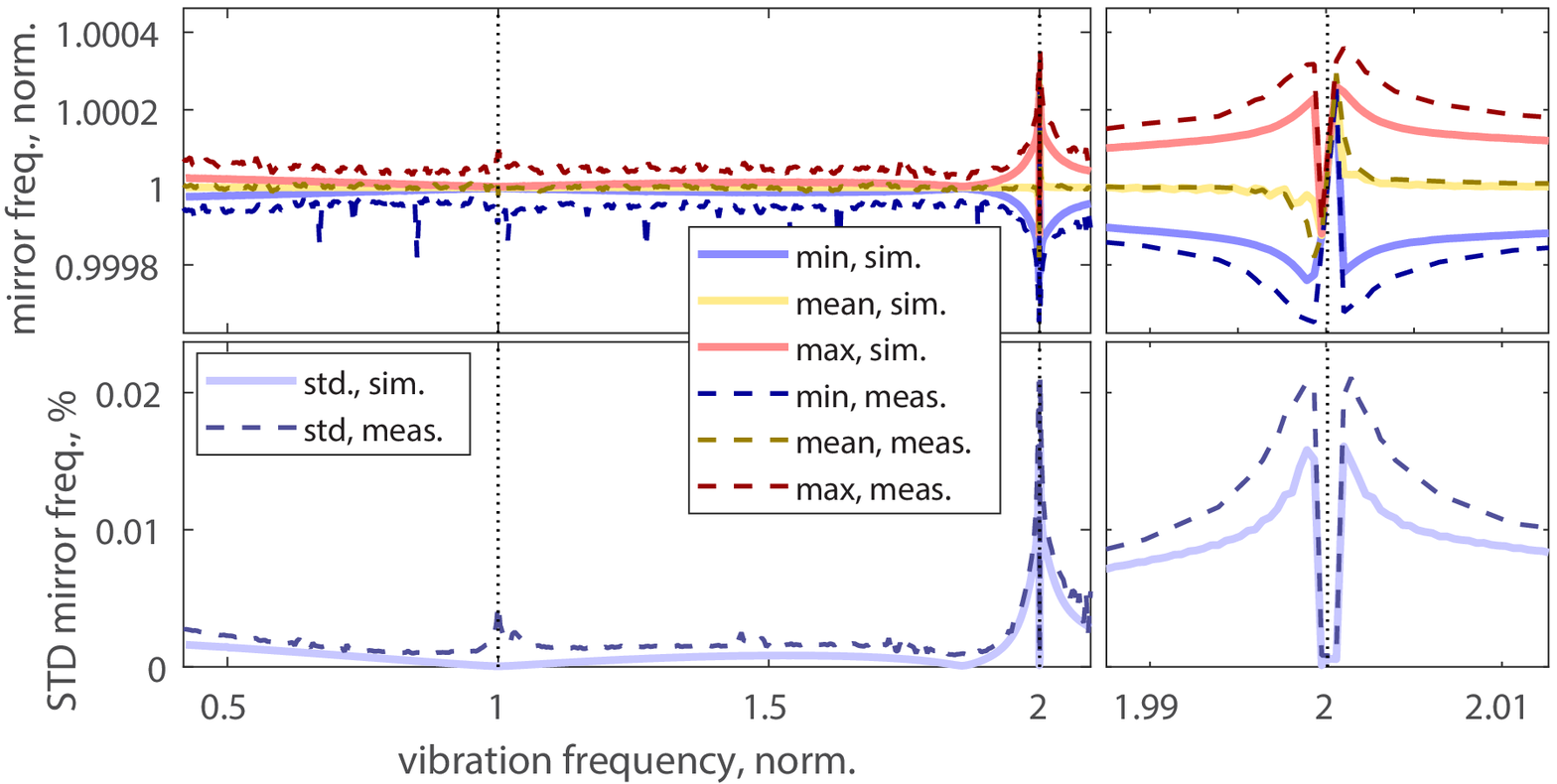}} 
	\caption[]{Tz vibration influence on the amplitude and  frequency for the MEMS  mirror with a PLL. The simulation and measurement are drawn with solid lines with light colors and dashed lines with dark colors, respectively. } 
\label{fig:TzSweepPLL}
\end{figure}
The vibration influence on the PLL-controlled mirror is also evaluated by both transient response and the vibration frequency sweep.
 Fig.~\ref{fig:TyTransientPLL} shows a transient vibration response of the PLL-controlled mirror for Ty vibration with a frequency of 1.0013. 
Contrary to the open loop case, the  mirror frequency shifts to the vibration frequency and the amplitude also follows a similar way to a high amplitude of 1.009. %
Transient vibration response for Tz vibration with the PLL is omitted since it is analog to the Ty PLL vibration case, however with the major influence around the mirror actuation frequency as in the open loop case.

The vibration frequency sweep in Fig.~\ref{fig:TySweepPLL} shows the behavior of mirror amplitude and frequency for the PLL-controlled mirror. 
For the Ty vibration with small frequency differences within $\pm 0.0029$, the PLL follows the vibration frequency and the amplitude also changes accordingly with frequency in a linear manner. In this region, there is no amplitude and frequency oscillation observed. 
For vibrations with a larger frequency difference over $\pm 0.0029$, amplitude and frequency oscillates as in the open loop case 
with the frequency of the maximum influence at $\pm$0.0029 relative to the mirror frequency.
The STD of the amplitude and frequency oscillation are 0.91~\%  and 0.15~\%, respectively, which is 44.4~\% and 43.0~\% reduction, respectively, compared to the open loop case. 
For a MEMS mirror with a 15$^\circ$ amplitude and a 2~kHz oscillation, these STD amplitude and frequency errors correspond to 0.137$^\circ$ and 3~Hz, respectively.

Fig.~\ref{fig:TzSweepPLL} illustrates the vibration frequency sweep for Tz vibration with the PLL enabled.
As for Ty vibration, a vibration with a small frequency difference makes the PLL follow the vibration, where the range is only $\pm 0.0007$ relative to the mirror actuation frequency.  %
A higher frequency difference outside of $\pm 0.0007$ relative to the mirror actuation frequency results in STD mirror amplitude and frequency errors of up to 0.13~\% and 0.02 ~\%, respectively. 
The improvement compared to the open loop system corresponds to 33.5~\% and 27.6~\% for STD amplitude and frequency errors, respectively.  

\subsection{Discussion}
The mismatch between the center of mass and the rotation axis is identified as the main cause of the vibration influence on the resonant MEMS system, supported by a good agreement of the various measurements and the simulations. %
The simulation also tracks asymmetries of the vibration influences for positive and negative frequency differences in the open loop case. %
The measurements verify the four aspects in the analysis in Sec.~\ref{se:energyAnalysis} as well. 

Since the vibration coupling mainly occurs at vibration frequencies near $f_\textrm{m}$ and $2f_\textrm{m}$, 
the mirror frequency can be chosen high enough by design to be beyond the band in which significant vibration occur in a specific application scenario.
For example, many standards show frequencies beyond 2 kHz as uncritical  \cite{VW8000Std,grahmannVibrationAnalysisMicro2020}, which can be used for the mirror scanning frequency.
Due to the directional dependency and the narrow frequency range of the vibration sensitivity, a design of a vibration isolation packages for MEMS lidar systems can be simplified by suppressing the dominant Ty vibration near the mirror frequency.

Beside the Ty and Tz vibrations, Tx vibrations are neglected because they do not directly couple to the Rx motion. 
Tx vibrations couple only to the orthogonal rotational mode of Ry instead of the operational mirror mode Rx by a mismatch between the center of mass and the rotation axis. 
Therefore, this coupling does not appear in the SDoF model (\ref{eq:ParamDuffOsci}). 
If all rotational DoF (Rx, Ry, and Rz) are included in the model, coupling between those rotational modes, mediated by Euler's equations, would arise \cite{frangiModeCouplingParametric2018}. This would also lead to an indirect coupling of Tx vibrations to the Rx mode. 
However, by design of the MEMS mirror, the parasitic Ry mode is significantly stiffer than the Rx mode with an almost 20 times higher eigenfrequency, as calculated by finite element analysis. 
For this reason, the sketched mechanism is expected to be far weaker than the one proposed for Ty and Tz vibrations. 
Experimental results with Tx vibrations hardly show any significant coupling influence on the mirror scanning trajectories and therefore they are omitted for simplicity.

The developed MEMS vibration test setup can evaluate the influence of translational vibrations for open loop and PLL-controlled MEMS mirrors, verifying the proposed vibration coupling mechanism of the MEMS mirror. 
The PLL also exhibits a reduction of vibration coupling influence by 43.0~\% in STD frequency errors compared to that of the open loop operation, showing the strong benefits of a controlled MEMS scanning system for reliable automotive MEMS lidars. 
\section{Conclusion}
This paper discusses an evaluation of vibration immunity of a resonant MEMS scanning system for automotive lidar applications. 
The MEMS mirror has a reinforcement structure on the backside of the mirror to reduce dynamic deformation, leading to a vibration-induced torque caused by the mismatch between the center of mass and the rotation axis.
The vibration-induced torque is analyzed by energy variation per period, showing directional dependency of vibration coupling.  %
A vibration test setup is developed to evaluate transient vibration response and vibration frequency sweep using 2~$g_\textrm{rms}$ single tone vibrations. 
The measurement results verify the analysis of vibration coupling by a good agreement with simulations, illustrating the directional dependency of vibration coupling.
A vibration frequency sweep  shows that the STD amplitude and frequency are up to 1.64~\% and 0.26~\%, respectively, for Ty vibrations.   
The PLL-controlled mirror improves the STD amplitude and frequency to 0.91~\% and 0.15~\%, respectively, which corresponds to 44.4~\%  and 43.0~\%  reduction in STD amplitude and frequency, respectively. 
Comparisons from these experimental investigations for worst case vibration excitations demonstrate that the proposed vibration test setup can accurately evaluate vibration coupling of the resonant MEMS scanning system in operation. 
The proposed vibration coupling model also allows for thorough investigation of vibration sensitivity of resonant MEMS mirrors even in the design phase.

\bibliographystyle{IEEEtranTIE}
\bibliography{./bibliography_reduced}   

\begin{thebibliography}{10}
\providecommand{\url}[1]{#1}
\csname url@samestyle\endcsname
\providecommand{\newblock}{\relax}
\providecommand{\bibinfo}[2]{#2}
\providecommand{\BIBentrySTDinterwordspacing}{\spaceskip=0pt\relax}
\providecommand{\BIBentryALTinterwordstretchfactor}{4}
\providecommand{\BIBentryALTinterwordspacing}{\spaceskip=\fontdimen2\font plus
\BIBentryALTinterwordstretchfactor\fontdimen3\font minus
  \fontdimen4\font\relax}
\providecommand{\BIBforeignlanguage}[2]{{%
\expandafter\ifx\csname l@#1\endcsname\relax
\typeout{** WARNING: IEEEtran.bst: No hyphenation pattern has been}%
\typeout{** loaded for the language `#1'. Using the pattern for}%
\typeout{** the default language instead.}%
\else
\language=\csname l@#1\endcsname
\fi
#2}}
\providecommand{\BIBdecl}{\relax}
\BIBdecl

\bibitem{royoOverviewLidarImaging2019}
S.~Royo and M.~Ballesta-Garcia, ``An {Overview} of {Lidar} {Imaging} {Systems}
  for {Autonomous} {Vehicles},'' \emph{Applied Sciences}, vol.~9, no.~19, p.
  4093, Jan. 2019.

\bibitem{Yoo18}
H.~W. Yoo, N.~Druml, D.~Brunner, C.~Schwarzl, T.~Thurner, M.~Hennecke, and
  G.~Schitter, ``{MEMS}-based lidar for autonomous driving,''
  \emph{Elektrotechnik und Informationstechnik}, vol. 135, no.~6, pp. 408--418,
  2018.

\bibitem{druml_1d_2018}
N.~Druml, I.~Maksymova, T.~Thurner, D.~van Lierop, M.~Hennecke, and
  A.~Foroutan, ``{1D} {MEMS} {Micro}-{Scanning} {LiDAR},'' in \emph{Int. Conf.
  on Sensor Device Technologies and Appl.}, 2018.

\bibitem{Kasturi2020Comp}
A.~Kasturi, V.~Milanovi\`{c}, D.~Lovell, F.~Hu, D.~Ho, Y.~Su, and L.~Ristic,
  ``{Comparison of MEMS mirror LiDAR architectures},'' in \emph{Proc. SPIE
  11293}, pp. 63 -- 79, 2020.

\bibitem{Ito13}
K.~Ito, C.~Niclass, I.~Aoyagi, H.~Matsubara, M.~Soga, S.~Kato, M.~Maeda, and
  M.~Kagami, ``System design and performance characterization of a mems-based
  laser scanning time-of-flight sensor based on a 256x64-pixel single-photon
  imager,'' \emph{IEEE Photonics J.}, vol.~5, p. 6800114, 2013.

\bibitem{hofmannResonantBiaxial7mm2013}
U.~Hofmann, M.~Aikio, J.~Janes, F.~Senger, V.~Stenchly, J.~Hagge, H.-J.
  Quenzer, M.~Weiss, T.~v. Wantoch, C.~Mallas, B.~Wagner, and W.~Benecke,
  ``Resonant biaxial 7-mm {MEMS} mirror for omnidirectional scanning,''
  \emph{J. Micro/Nanolith. MEMS MOEMS}, vol.~13, no.~1, p. 011103, Dec. 2013.

\bibitem{Kim16}
G.~Kim and Y.~Park, ``Lidar pulse coding for high resolution range imaging at
  improved refresh rate,'' \emph{Opt. Express}, vol.~24, pp. 23\,810--23\,828,
  2016.

\bibitem{Tha16}
R.~Thakur, ``Scanning lidar in advanced driver assistance systems and beyond,''
  \emph{IEEE Consumer Electronics Magazine}, pp. 48--54, 2016.

\bibitem{Sta16}
B.~L. Stann, J.~F. Dammann, and M.~M. Giza, ``Progress on mems-scanned ladar,''
  in \emph{Proc. SPIE 9832}, p. 98320L, Jul. 2016.

\bibitem{sandnerHybridAssembledMEMS2019}
T.~Sandner, T.~Grasshoff, M.~Wildenhain, and M.~Schwarzenberg, ``Hybrid
  assembled {MEMS} scanner array with large aperture for fast scanning {LIDAR}
  systems,'' \emph{tm - Technisches Messen}, vol.~86, no.~3, pp. 151--163,
  2019.

\bibitem{wolter_applications_2005-1}
A.~Wolter, S.-T. Hsu, H.~Schenk, and H.~K. Lakner, ``Applications and
  requirements for {MEMS} scanner mirrors,'' in \emph{Proc. SPIE 5719}, pp.
  64--76, Jan. 2005.

\bibitem{VW8000Std}
``\BIBforeignlanguage{en}{{LV 124}: Electric and electronic components in motor
  vehicles up to 3.5 t - general requirements, test conditions and tests ({VW
  80000})},'' Volkswagen, Tech. Rep. 8MA00, 2013.

\bibitem{yooMEMSMicromirrorCharacterization2009}
B.-W. Yoo, J.-H. Park, I.~H. Park, J.~Lee, M.~Kim, J.-Y. Jin, J.-A. Jeon, S.-W.
  Kim, and Y.-K. Kim, ``{MEMS} micromirror characterization in space
  environments,'' \emph{Opt. Express, OE}, vol.~17, no.~5, pp. 3370--3380, Mar.
  2009.

\bibitem{VerticalCombs}
D.~Hah, P.~R. Patterson, H.~D. Nguyen, H.~Toshiyoshi, and M.~C. W, ``{Theory
  and Experiments of Angular Vertical Comb-Drive Actuators for Scanning
  Micromirrors},'' \emph{IEEE J. Sel. Top. Quantum Electron.}, vol.~10, no.~3,
  2004.

\bibitem{Jung2012QSFraunhoferIPMS}
D.~Jung, T.~Sandner, D.~Kallweit, and H.~Schenk, ``{Vertical comb drive
  microscanners for beam steering, linear scanning, and laser projection
  applications},'' in \emph{Proc. SPIE 8252}, pp. 232 -- 241, 2012.

\bibitem{hofmann_high-q_2012}
U.~Hofmann, J.~Janes, and H.-J. Quenzer, ``High-{Q} {MEMS} {Resonators} for
  {Laser} {Beam} {Scanning} {Displays},'' \emph{Micromachines}, vol.~3, no.~2,
  pp. 509--528, 2012.

\bibitem{milanovicNovelFluidicPackaging2016}
V.~Milanovi\`{c}, A.~Kasturi, and J.~Yang, ``Novel fluidic packaging of
  gimbal-less {MEMS} mirrors for increased optical resolution and overall
  performance,'' in \emph{Proc. SPIE 9836}, p. 98362Z, May. 2016.

\bibitem{milanovicIterativeLearningControl2018}
V.~Milanovi\`{c}, A.~Kasturi, H.~J. Kim, and F.~Hu, ``Iterative learning
  control ({ILC}) algorithm for greatly increased bandwidth and linearity of
  {MEMS} mirrors in {LiDAR} and related imaging applications,'' in \emph{Proc.
  SPIE 10545}, p. 1054513, Feb. 2018.

\bibitem{leiFR4BasedElectromagneticScanning2018}
H.~Lei, Q.~Wen, F.~Yu, Y.~Zhou, and Z.~Wen, ``{FR4}-{Based} {Electromagnetic}
  {Scanning} {Micromirror} {Integrated} with {Angle} {Sensor},''
  \emph{Micromachines}, vol.~9, no.~5, p. 214, May. 2018.

\bibitem{grahmannVibrationAnalysisMicro2020}
J.~Grahmann, R.~Schroedter, O.~Kiethe, and U.~Todt, ``Vibration analysis of
  micro mirrors for {LIDAR} using on-chip piezo-resistive sensor,'' in
  \emph{Proc SPIE 11293}, p. 1129308, Feb. 2020.

\bibitem{urey_optical_2000}
H.~Urey, D.~W. Wine, and T.~D. Osborn, ``\BIBforeignlanguage{en}{Optical
  performance requirements for {MEMS}-scanner-based microdisplays},'' in
  \emph{\BIBforeignlanguage{en}{Proc. SPIE vol. 4178}}, pp. 176--185, Aug.
  2000.

\bibitem{sandner_synchronized_2010}
T.~Sandner, T.~Grasshoff, M.~Wildenhain, and H.~Schenk, ``Synchronized
  microscanner array for large aperture receiver optics of {LIDAR} systems,''
  in \emph{{MOEMS} and {Miniaturized} {Systems} {IX}}, vol. 7594, p.
  75940C.\hskip 1em plus 0.5em minus 0.4em\relax International Society for
  Optics and Photonics, Feb. 2010.

\bibitem{neeLightweightOpticallyFlat2000}
J.~Nee, R.~Conant, R.~Muller, and K.~Lau, ``Lightweight, optically flat
  micromirrors for fast beam steering,'' in \emph{2000 {IEEE}/{LEOS} {Int.}
  {Conf.} on {Opt.} {MEMS}}, pp. 9--10, Aug. 2000.

\bibitem{milanovicGimballessMonolithicSilicon2004}
V.~Milanovic, G.~Matus, and D.~McCormick, ``Gimbal-less monolithic silicon
  actuators for tip-tilt-piston micromirror applications,'' \emph{IEEE J. Sel.
  Topics Quantum Electron.}, vol.~10, no.~3, pp. 462--471, May. 2004.

\bibitem{hsuFabricationCharacterizationDynamically2008}
S.~Hsu, T.~Klose, C.~Drabe, and H.~Schenk, ``Fabrication and characterization
  of a dynamically flat high resolution micro-scanner,'' \emph{J. Opt. A: Pure
  Appl. Opt.}, vol.~10, no.~4, p. 044005, 2008.

\bibitem{YooCTF2020}
H.~W. Yoo and G.~Schitter, ``Complex valued state space model for weakly
  nonlinear {Duffing} oscillator with noncollocated external disturbance,'' in
  \emph{21st IFAC World Congress}, p. in press, Jul. 2020.

\bibitem{leonardusMEMSSCANNINGMICROMIRROR2018}
H.~W. L. A.~M. van Lierop and M.~A.~G. Suijlen, ``{MEMS} scanning
  micromirror,'' US Patent 9,910,269 B2, Mar., 2018.

\bibitem{DavidSPIE19}
D.~Brunner, H.~W. Yoo, T.~Thurner, and G.~Schitter, ``Data based modelling and
  identification of nonlinear {SDOF} {MOEMS} mirror,'' in \emph{Proc. SPIE
  10931}, p. 1093117, 2019.

\bibitem{YooStephanMEMSDFT20}
H.~W. Yoo, S.~Albert, and G.~Schitter, ``Accurate analytic model of a
  parametrically driven resonant {MEMS} mirror with a {Fourier} series based
  torque approximation,'' \emph{J. Microelectromech. Syst.}, vol.~5, no.~29,
  pp. 1431--1442, 2020.

\bibitem{Ataman2006}
C.~Ataman and H.~Urey, ``Modeling and characterization of comb-actuated
  resonant microscanners,'' \emph{J. Micromech. Microeng.}, vol.~16, no.~1,
  p.~9, 2006.

\bibitem{frangiParametricResonanceElectrostatically2017}
A.~Frangi, A.~Guerrieri, R.~Carminati, and G.~Mendicino, ``Parametric
  {Resonance} in {Electrostatically} {Actuated} {Micromirrors},'' \emph{IEEE
  Trans. Ind. Electron.}, vol.~64, no.~2, pp. 1544--1551, Feb. 2017.

\bibitem{frangiAccurateSimulationParametrically2017}
A.~Frangi, A.~Guerrieri, and N.~Boni, ``Accurate {Simulation} of
  {Parametrically} {Excited} {Micromirrors} via {Direct} {Computation} of the
  {Electrostatic} {Stiffness},'' \emph{Sensors}, vol.~17, no.~4, p. 779, 2017.

\bibitem{kovacicMathieuEquationIts2018}
I.~Kovacic, R.~Rand, and S.~Mohamed~Sah, ``\BIBforeignlanguage{en}{Mathieu's
  {Equation} and {Its} {Generalizations}: {Overview} of {Stability} {Charts}
  and {Their} {Features}},'' \emph{\BIBforeignlanguage{en}{Appl. Mech. Rev.}},
  vol.~70, no.~2, p. 020802, Mar. 2018.

\bibitem{DavidTIE20}
D.~Brunner, H.~W. Yoo, and G.~Schitter, ``Linear modeling and control of
  comb-actuated resonant {MEMS} mirror with nonlinear dynamics,'' \emph{IEEE
  Trans. Ind. Electron.}, vol.~67, no.~-, p. In press, 2020.

\bibitem{YooTestBench2019}
H.~W. Yoo, D.~Brunner, T.~Thurner, and G.~Schitter, ``{MEMS} test bench and its
  uncertainty analysis for evaluation of {MEMS} mirrors,'' in \emph{8th IFAC
  Symp. on Mechatronic Syst.}, pp. 49--54, Sep. 2019.

\bibitem{Ievgeniia2019SPIE}
I.~Maksymova, P.~Greiner, J.~Wiesmeier, F.~Darrer, and N.~Druml, ``A mems
  mirror driver asic for beam-steering in scanning mems-based lidar,'' in
  \emph{Proc. SPIE 11107}, p. 111070C, Sep. 2019.

\bibitem{frangiModeCouplingParametric2018}
A.~Frangi, A.~Guerrieri, N.~Boni, R.~Carminati, M.~Soldo, and G.~Mendicino,
  ``Mode {Coupling} and {Parametric} {Resonance} in {Electrostatically}
  {Actuated} {Micromirrors},'' \emph{IEEE Trans. on Ind. Electron.}, vol.~65,
  no.~7, pp. 5962--5969, Jul. 2018.

\end{thebibliography}

\end{document}